\documentclass[aps,prb,twocolumn]{revtex4-1}

\usepackage[pdftex]{graphics}
\usepackage{amssymb,amsmath}
\usepackage{nicefrac}
\usepackage{color}

\begin{document}
\def\bra#1{\left<{#1}\right|}
\def\ket#1{\left|{#1}\right>}
\def\expval#1#2{\bra{#2} {#1} \ket{#2}}
\def\mapright#1{\smash{\mathop{\longrightarrow}\limits^{_{_{\phantom{X}}}{#1}_{_{\phantom{X}}}}}}

\title{Spin-dependent charge recombination along para-phenylene molecular wires}
\author{Thomas P.~Fay}
\affiliation{Department of Chemistry, University of Oxford, Physical and Theoretical Chemistry Laboratory, South Parks Road, Oxford, OX1 3QZ, UK}
\author{Alan M.~Lewis}
\affiliation{Department of Chemistry, University of Oxford, Physical and Theoretical Chemistry Laboratory, South Parks Road, Oxford, OX1 3QZ, UK}
\author{David E. Manolopoulos}
\affiliation{Department of Chemistry, University of Oxford, Physical and Theoretical Chemistry Laboratory, South Parks Road, Oxford, OX1 3QZ, UK}

\begin{abstract}
We have used an efficient new quantum mechanical method for radical pair recombination reactions to study the spin-dependent charge recombination along PTZ$^{\bullet+}$--Ph$_n$--PDI$^{\bullet-}$ molecular wires. By comparing our results to the experimental data of E. Weiss {\em et al.} [J. Am. Chem. Soc. {\bf 126}, 5577 (2004)], we are able to extract the spin-dependent (singlet and triplet) charge recombination rate constants for wires with $n=2-5$. These spin-dependent rate constants have not been extracted previously from the experimental data because they require fitting its magnetic field-dependence to the results of quantum spin dynamics simulations. We find that the triplet recombination rate constant decreases exponentially with the length of the wire, consistent with the superexchange mechanism of charge recombination. However, the singlet recombination rate constant is nearly independent of the length of the wire, suggesting that the singlet pathway is dominated by an incoherent hopping mechanism. A simple qualitative explanation for the different behaviours of the two spin-selective charge recombination pathways is provided in terms of Marcus theory. We also find evidence for a magnetic field-independent background contribution to the triplet yield of the charge recombination reaction, and suggest several possible explanations for it. Since none of these explanations is especially compelling given the available experimental evidence, and since the result appears to apply more generally to other molecular wires, we hope that this aspect of our study will stimulate further experimental work.
\end{abstract}

\maketitle

\section{Introduction}

\begin{figure}[t]
\centering
\resizebox{0.9\columnwidth}{!} {\includegraphics{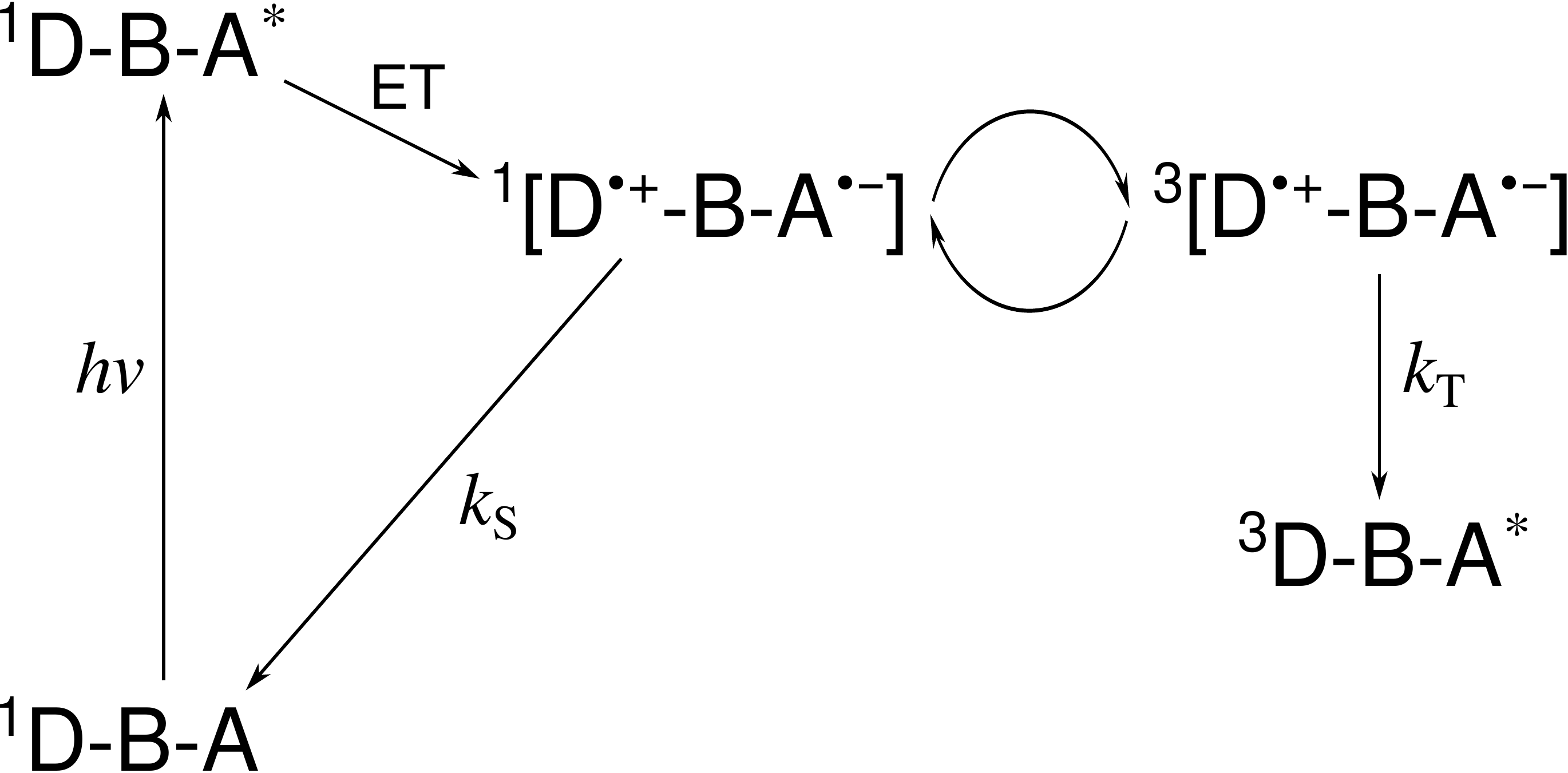}}
\caption{A typical charge recombination reaction in a donor-bridge-acceptor (D-B-A) molecular wire. The semicircular arrows indicate hyperfine mediated intersystem crossing between the singlet and triplet states of the charge separated D$^{\bullet+}$-B-A$^{\bullet-}$ radical pair, which is formed by photoexcitation of a singlet D-B-A precursor followed by electron transfer. The focus of the present study is on extracting the singlet and triplet charge recombination rate constants $k_{\rm S}$ and $k_{\rm T}$ from the magnetic field dependence of experimental observables such as the triplet yield of the charge recombination reaction.}
\label{RPR}
\end{figure}

In recent years, molecular wires have been the subject of significant interest and investigation.\cite{Weiss05b,Goldsmith05} They are designed to mimic the efficient long range charge transport found in the photosynthetic reaction centre, and have a range of possible applications.\cite{Parson03} In particular, efficient `wire-like' charge separation is a highly desirable feature in chemical solar energy conversion systems.\cite{Wasielewski06} In order to design suitable molecular wires, an understanding of the mechanisms by which unproductive charge recombination can occur along them is also clearly desirable.\cite{Jortner97}

A typical charge recombination process in an organic donor-bridge-acceptor (D-B-A) molecular wire is shown in Fig.~1. The initial charge separated D$^{\bullet+}$-B-A$^{\bullet-}$ radical pair is formed in its singlet state by photoexcitation of the ground state $^1$D-B-A molecule and subsequent electron transfer along the wire. This then interconverts with the triplet state via hyperfine-mediated intersystem crossing, which competes with the depletion of the radical pair by spin-dependent charge recombination reactions leading to singlet and triplet products (with rate constants $k_{\rm S}$ and $k_{\rm T}$, respectively).

There are two limiting mechanisms of charge recombination along the wire: the superexchange mechanism and the incoherent hopping mechanism. In the former, recombination occurs by an electron tunnelling from the A$^{\bullet-}$ radical to the D$^{\bullet+}$ radical in a single step via superexchange coupling between the electron and hole, which is mediated by orbitals on the bridge.\cite{Anderson59} The magnitude of this coupling, and therefore the rate of electron transfer by this mechanism, is expected to decrease exponentially with increasing donor-acceptor separation.\cite{McConnell61,Jortner98} By contrast, the incoherent hopping mechanism is a two step process, in which an electron hops from the bridge to the D$^{\bullet+}$ radical, followed by a second electron hopping from the A$^{\bullet-}$ radical onto the bridge. The rate of charge transfer by this mechanism is governed by the energy gap between the charge separated state of the molecular wire, D$^{\bullet+}$--B--A$^{\bullet-}$, and the intermediate formed in the first electron hopping step, D--B$^{\bullet+}$--A$^{\bullet-}$. Provided this energy gap is small, the rate of electron transfer is approximately independent of the radical pair separation.\cite{Berlin02}

In general, the charge recombination along a wire in the singlet state will occur at a different rate, and potentially by a different mechanism, than a wire in the triplet state. The hyperfine-mediated intersystem crossing between these spin states can therefore play a significant role in the overall rate of charge recombination. Since this intersystem crossing is affected by an applied magnetic field,\cite{Steiner89} so too is the rate of charge recombination, and the magnetic field dependence of experimental observables such as the triplet yield of the recombination reaction can be used to shed light on the rates and mechanisms of the singlet and triplet recombination pathways. However, to do this properly requires fitting the experimental data to quantum mechanical spin dynamics simulations. While this is straightforward to do for small D$^{\bullet+}$ and A$^{\bullet-}$ radicals, it can become computationally expensive for radicals with many hyperfine-coupled nuclear spins.

In a recent paper,\cite{Lewis16} we have developed an efficient method for solving this problem, based on a Monte Carlo evaluation of the triplet yield in an overcomplete basis of nuclear spin coherent states (see Sec.~II). This new method is routinely applicable to radical pairs with as many as 20 or so hyperfine-coupled nuclear spins, which is the typical size of radical pair encountered in experimental studies of molecular wires.\cite{Weiss05b,Goldsmith05,Hasharoni95,Boom02,Weiss03,Weiss04,Weiss05,Dance06} It has also been shown to be more reliable than the various semiclassical approximations\cite{Schulten78,Manolopoulos13,Lewis14,Lawrence16} that can be applied to the problem.\cite{Lewis16} In this paper, we shall exploit our new method by applying it to the PTZ$^{\bullet+}$--Ph$_n$--PDI$^{\bullet-}$ molecular wires studied by Weiss {\em et al.},\cite{Weiss04} which consist of a phenothiazine (PTZ) donor, a perylene-3,4:9,10-bis(dicarboximide) (PDI) acceptor, and a bridge of $n$ para-phenylene rings.

In their experimental study, Weiss {\em et al.}\cite{Weiss04} used transient absorption spectroscopy to measure both the charge separation and (overall) charge recombination rate constants for the wires with $n=1-5$, and found evidence for a change in the mechanisms of both processes as the length of the bridge was increased. For wires with short bridges, both rate constants were found to decrease exponentially with increasing bridge length, consistent with the superexchange mechanism. However, for the longest wires considered in the experiments, the rate constants were found to increase slightly with increasing $n$ (beyond $n=4$ in the case of charge separation and $n=3$ in the case of charge recombination), which was taken to indicate a change in mechanism to incoherent hopping.\cite{Weiss04}

The overall charge recombination rate constant measured by transient absorption spectroscopy of PTZ$^{\bullet+}$--Ph$_n$--PDI$^{\bullet-}$ contains contributions from both the singlet and triplet charge recombination pathways shown in Fig.~1, but the relative contributions of these two pathways were not extracted from the experimental data.\cite{Weiss04} Here we shall show how this can be done by fitting the measured magnetic field dependence of the triplet yield, and of the radical pair survival probability 50 ns after the initial photoexcitation pulse, to the results of quantum spin dynamics simulations. In doing so, we shall be able to disentangle the contributions of the singlet and triplet pathways to the overall charge recombination rate, and reveal the likely mechanisms by which these pathways operate.

\section{Theory}

In order to simulate the charge recombination of the charge separated D$^{\bullet+}$-B-A$^{\bullet-}$ radical pair, we must first define a Hamiltonian under which the electron spins in the two radicals evolve, and a recombination operator that accounts for the singlet and triplet charge recombination processes. 

For a molecular wire that is tumbling in solution, the spin Hamiltonian contains isotropic Zeeman and hyperfine interactions, and an exchange interaction between the electron spins:\cite{Wasielewski06}
\begin{equation}
\begin{gathered}
\hat{H} = \hat{H}_1 + \hat{H}_2 + 2J\,\hat{\bf S}_1\cdot\hat{\bf S}_2, \\
\hat{H}_i = \boldsymbol{\omega}_i\cdot\hat{\bf S}_i+\sum_{k=1}^{N_i} a_{ik}\,\hat{\bf I}_{ik}\cdot\hat{\bf S}_i.
\label{Hamiltonian}
\end{gathered}
\end{equation}
Here $\boldsymbol{\omega}_i = -\gamma_i{\bf B}$, where $\gamma_i$ is the gyromagnetic ratio of the electron in radical $i$ and ${\bf B}$ is the applied magnetic field. In each radical, $a_{ik}$ is the hyperfine coupling constant between the electron spin and the $k$th nuclear spin, $\hat{\bf S}_i$ and $\hat{\bf I}_{ik}$ are the corresponding electron and nuclear spin angular momentum operators, and $N_i$ is the total number of hyperfine-coupled nuclear spins. $J$ is the exchange coupling between the electron spins. Note that we have neglected the comparatively weak Zeeman interactions of the nuclear spins with the applied magnetic field, and that we are working in a unit system in which $\hbar=1$.

The singlet and triplet charge recombination processes can be modelled using the Haberkorn recombination operator,\cite{Haberkorn76,Ivanov10}
\begin{equation}
\hat{K} = {k_{\rm S}\over 2}\hat{P}_{\rm S}+{k_{\rm T}\over 2}\hat{P}_{\rm T},
\label{Haberkorn}
\end{equation}
where
\begin{equation}
\hat{P}_{\rm S} = {1\over 4}\hat{\bf 1}-\hat{\bf S}_1\cdot\hat{\bf S}_2
\end{equation}
and
\begin{equation}
\hat{P}_{\rm T} = {3\over 4}\hat{\bf 1}+\hat{\bf S}_1\cdot\hat{\bf S}_2
\label{P_T_op}
\end{equation}
are the projection operators onto the electronic singlet and triplet states of the radical pair and $k_{\rm S}$ and $k_{\rm T}$ are the corresponding first order charge recombination rate constants. 

The spin dynamics of the PTZ$^{\bullet+}$--Ph$_n$--PDI$^{\bullet-}$ molecular wires has been probed experimentally by measuring the effect of an applied magnetic field on both the triplet yield and what Weiss {\em et al.}\cite{Weiss04} term the \lq\lq radical pair yield" -- the survival probability of the radical pair 50 ns after the initial photoexcitation laser pulse. The triplet yield is defined as\cite{Lewis14}
\begin{equation}
\Phi_{\rm T} = k_{\rm T} \int_0^{\infty} {\rm P_T}(t)\, {\rm d}t,
\label{tyield}
\end{equation}
where ${\rm P_T}(t)$ is the ensemble average of the triplet probability,
\begin{equation}
{\rm P_T}(t) = {\rm Tr}[\hat{\rho}(t)\hat{P}_{\rm T}],
\label{P_T}
\end{equation}
and 
\begin{equation}
\hat{\rho}(t) = e^{-i\hat{H}t-\hat{K}t}\hat{\rho}(0)e^{+i\hat{H}t-\hat{K}t}
\end{equation}
is the density operator of the spin system at time $t$. Since the charge separated radical pair is formed in its singlet state, its initial density operator is 
\begin{equation}
\hat{\rho}(0)=\frac{1}{Z}\hat{P}_{\rm S},
\end{equation}
where $Z=\prod_{i=1}^2\prod_{k=1}^{N_i} I_{ik}(I_{ik}+1)$ is the total number of nuclear spin states in the two radicals. The radical pair yield is simply
\begin{equation}
\Phi_{\rm RP} = {\rm Tr}[\hat{\rho}(t)],
\label{1}
\end{equation}
evaluated at $t \simeq 50$ ns. 

The standard way to evaluate the traces in Eqs.~(6) and~(9) is to exploit the structure of $\hat{\rho}(0)$ in Eq.~(8) and to write each trace as a sum over $Z$ initial nuclear spin projection states. However, this leads to $Z$ independent time-dependent wavepacket propagations, which is an enormous number for a radical pair with 20 or so nuclear spins. It is considerably more efficient to proceed by letting\cite{Lewis16}
\begin{equation}
\ket{{\rm S},{\bf \Omega}_1,{\bf \Omega}_2;t} = e^{-i\hat{H}t-\hat{K}t} \ket{{\rm S},{\bf \Omega}_1,{\bf \Omega}_2},
\label{prop}
\end{equation}
where $\ket{\rm S}$ is the singlet electronic spin state,
\begin{equation}
\ket{{\bf \Omega}_i} = \ket{\Omega_{i1}} \otimes \ket{\Omega_{i2}} \otimes \dots \otimes \ket{\Omega_{iN_i}},
\end{equation}
and $\ket{\Omega_{ik}}$ is a coherent spin state\cite{Radcliffe71} of the $k$th nuclear spin in radical $i$. Then Eq.~(5) becomes\cite{Lewis16}
\begin{equation}
\begin{aligned}
\Phi_{\rm T} &= \frac{1}{(4\pi)^N}\int {\rm d}{\bf \Omega}_1 \int {\rm d}{\bf \Omega}_2 \\
                    &\times k_{\rm T}\int_0^{\infty}\expval {\hat{P}_{\rm T}}{{\rm S},{\bf \Omega}_1,{\bf \Omega}_2;t}dt,
\end{aligned}
\end{equation}
and Eq.~(9) becomes
\begin{equation}
\begin{aligned}
\Phi_{\rm RP} &= \frac{1}{(4\pi)^N}\int {\rm d}{\bf \Omega}_1 \int {\rm d}{\bf \Omega}_2 \\
&\times \left<{\rm S},{\bf \Omega}_1,{\bf \Omega}_2;t \right.\left|{\rm S},{\bf \Omega}_1,{\bf \Omega}_2;t\right>
\end{aligned}
\end{equation}
with $t\simeq 50$ ns, where $N=N_1+N_2$ is the total number of nuclear spins in the radical pair.

In these equations, the integrals over ${\bf \Omega}_1$ and ${\bf \Omega}_2$ are over the directions of the coherent spin states of each of the nuclear spins in the radical pair. These integrals are therefore high ($2N$) dimensional. However, the integrands of both integrals are probabilities, which are bounded between 0 and 1. This implies that their standard deviations are bounded by $1/2$. This is the ideal situation for Monte Carlo integration, which can be implemented simply by sampling each initial nuclear coherent spin state direction $\Omega_{ik}$ at random from the surface of a sphere.\cite{Lewis16}

In the test calculations reported in Ref.~\onlinecite{Lewis16}, for a model radical pair with 20 nuclear spins, this Monte Carlo integration was found to be significantly more efficient than a deterministic evaluation of the trace in the standard basis of nuclear spin projection states. The results were  converged to graphical accuracy with just $M=200$ Monte Carlo samples, whereas the total number of nuclear spin projection states in the radical pair was over a million.\cite{Lewis16} We have also found that 200 Monte Carlo samples are enough to give results converged to graphical accuracy in all of the calculations we shall report below.

One caveat we should make about this method is that Eqs.~(12) and~(13) neglect electron spin relaxation. For the present application, we do not feel that this is an issue. The charge recombination lifetimes of the PTZ$^{\bullet+}$-Ph$_n$-PDI$^{\bullet-}$ molecular wires that we shall consider are significantly less than a microsecond,\cite{Weiss04} and the radical pair yield in Eq.~(13) is evaluated just 50 ns after the initial photoexcitation laser pulse. It is unlikely that electron spin relaxation will have much effect over such a short timescale. For other applications, one might want to modify the theory to include electron spin relaxation. This can be done by coupling the spin dynamics to molecular motions that modulate the parameters in the spin Hamiltonian, as we shall show in a separate publication.\cite{Lindoy17} 

In the following section, we shall define the parameters that enter the Hamiltonian in Eq.~\eqref{Hamiltonian} and the recombination operator in Eq.~\eqref{Haberkorn} for the PTZ$^{\bullet+}$--Ph$_n$--PDI$^{\bullet-}$ molecular wires considered by Weiss {\em et al.}\cite{Weiss04} In Sec.~IV we  shall then move on to evaluate Eqs.~(12) and~(13) and calculate the triplet and radical pair yields of these molecular wires as a function of the strength of the applied magnetic field.

\section{Simulation Parameters}

\subsection{Spin Hamiltonian}

\begin{figure}[t]
\centering
\resizebox{0.7\columnwidth}{!} {\includegraphics{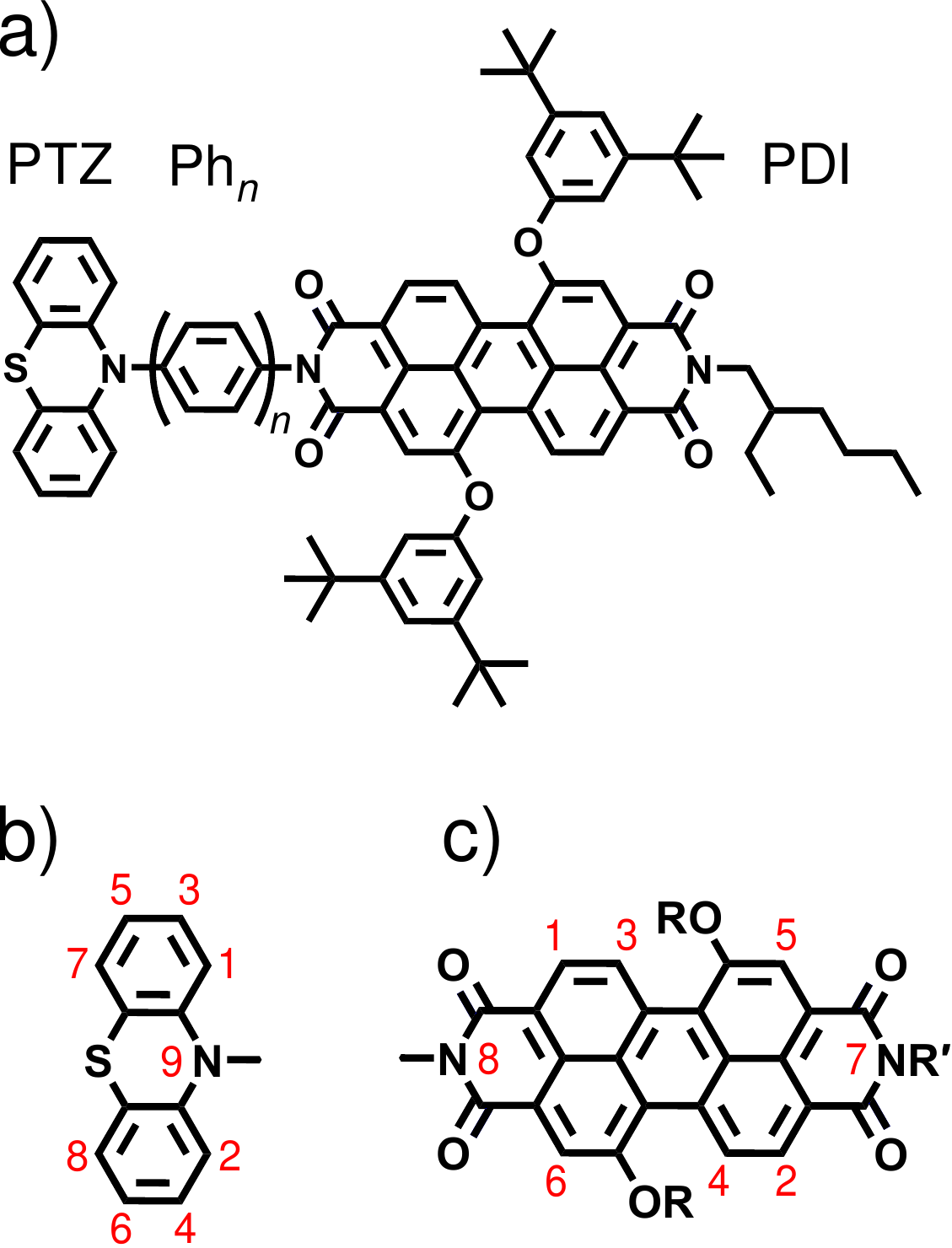}}
\caption{a) The chemical structure of a PTZ$^{\bullet+}$--Ph$_n$--PDI$^{\bullet-}$ molecular wire. b) The positions of the nuclei corresponding the hyperfine coupling constants of the PTZ$^{\bullet+}$ radical listed in Table \ref{PTZ_Hyps}. c) The positions of the nuclei corresponding the hyperfine coupling constants of the PDI$^{\bullet-}$ radical listed in Table \ref{PDI_Hyps}. }
\label{Chemical}
\end{figure}

The chemical structure of a PTZ$^{\bullet+}$--Ph$_n$--PDI$^{\bullet-}$ molecular wire is shown in Fig.~2. The spin evolution in this wire is governed by the interactions of the electron spins with the applied magnetic field, with the nuclear spins to which they are coupled, and with each other. We therefore need to specify the gyromagnetic ratios $\gamma_i$ of the two electrons, the hyperfine coupling constants in the two radicals, and the strength $J$ of the exchange coupling between the electron spins. 

For the gyromagnetic ratios, $\gamma_i=g_i\mu_{\rm B}/\hbar$, we have simply set $g_1=g_2=g_{\rm e}=2.0023$ in our simulations. This has the advantage that it eliminates the $\Delta g$ mechanism of intersystem crossing,\cite{Steiner89} which simplifies the interpretation of some of our results. The $g$ factors of the electrons in the isolated radicals have been measured by electron paramagnetic resonance (EPR) spectroscopy for PTZ$^{\bullet+}$ ($g_1=2.0053$),\cite{Chiu70} and by electron nuclear double resonance (ENDOR) spectroscopy for PDI$^{\bullet-}$ ($g_2=2.0028$).\cite{Tauber06} However, since $g$ factors can be quite sensitive to the substituents on a radical,\cite{Clarke77} it is not clear that they will be the same in PTZ$^{\bullet +}$--Ph$_n$--PDI$^{\bullet-}$. Also, the experiments of Weiss {\em et al.}\cite{Weiss04} were performed at relatively low magnetic field strengths ($B<120$ mT for the $n=3$-5 wires). The small differences between the $g$ factors of the electrons in the two radicals and that of a free electron will only have a tiny effect at these field strengths, and we have verified that using the isolated radical $g$ factors  does not change any of our results for $n=3$-5. For the $n=2$ wire, it does have a slight effect at the very highest field strength considered by Weiss {\em et al.}\cite{Weiss05}  ($B=600$ mT), as we shall discuss in Sec.~V.A. 

\begin{table} [b]
\centering
\begin{tabular} { l l l l }
\quad $k$\quad & \quad Nucleus\quad  & \quad $|a_{ik}|/|\gamma_{\rm e}|$ (Expt) \quad & \quad $a_{ik}/|\gamma_{\rm e}|$ (DFT) \quad \\ 
\hline
\hline
\quad 1 & \quad H & \quad 0.113 & \quad $-0.0753$ \\
\quad 2 & \quad H & \quad 0.113 & \quad $-0.0753$ \\
\quad 3 & \quad H & \quad 0.050 & \quad $-0.0813$ \\
\quad 4 & \quad H & \quad 0.050 & \quad $-0.0813$ \\
\quad 5 & \quad H & \quad 0.249 & \quad $-0.2247$ \\
\quad 6 & \quad H & \quad 0.249 & \quad $-0.2247$ \\
\quad 7 & \quad H & \quad 0.050 & \quad \phantom{$-$}0.0503 \\
\quad 8 & \quad H & \quad 0.050 & \quad \phantom{$-$}0.0503 \\
\quad 9 & \quad N & \quad 0.634 & \quad \phantom{$-$}0.3917 \\
\hline
\hline
\end{tabular}
\caption{The hyperfine coupling constants of the PTZ$^{\bullet +}$ radical in mT. Experimental data is taken from Ref.~\onlinecite{Chiu70}; DFT calculations were performed using the B3LYP functional with the cc-PV5Z basis set.}
\label{PTZ_Hyps}
\end{table}

\begin{table} [b]
\centering
\begin{tabular} { l l l l }
\quad $k$\quad & \quad Nucleus\quad & \quad $|a_{ik}|/|\gamma_{\rm e}|$ (Expt)\quad & \quad $a_{ik}/|\gamma_{\rm e}|$ (DFT)\quad \\
\hline
\hline
\quad 1 & \quad H & \quad 0.0785 & \quad \phantom{$-$}0.1351 \\
\quad 2 & \quad H & \quad 0.0785 & \quad \phantom{$-$}0.1351 \\
\quad 3 & \quad H & \quad 0.1720 & \quad $-0.2263$ \\
\quad 4 & \quad H & \quad 0.1720 & \quad $-0.2263$ \\
\quad 5 & \quad H & \quad 0.0575 & \quad \phantom{$-$}0.0658 \\
\quad 6 & \quad H & \quad 0.0575 & \quad \phantom{$-$}0.0658 \\
\quad 7 & \quad N & \quad 0.0621 & \quad $-0.0348$ \\
\quad 8 & \quad N & \quad 0.0621 & \quad $-0.0348$ \\
\hline
\hline
\end{tabular}
\caption{The hyperfine coupling constants of the PDI$^{\bullet -}$ radical in mT. Experimental data is taken from Ref.~\onlinecite{Tauber06}; DFT calculations were performed using the B3LYP functional with the EPR-II basis set.}
\label{PDI_Hyps}
\end{table}

The magnitudes of the hyperfine coupling constants $\{a_{ik}\}$ have also been measured by EPR for PTZ$^{\bullet+}$,\cite{Chiu70} and by ENDOR for PDI$^{\bullet-}$.\cite{Tauber06} The measured values are compared with the hyperfine couplings obtained from B3LYP\cite{Becke93,Lee88} density functional theory (DFT) calculations in Tables \ref{PTZ_Hyps} and \ref{PDI_Hyps}. For the purposes of these calculations, the O-R and N-R$'$ side chains in PDI$^{\bullet-}$ were replaced with O-H and N-H groups. 

The agreement between the experimental and calculated coupling constants is not especially good for either radical. This is highlighted by comparing the effective hyperfine fields,
\begin{equation}
B_{{\rm hyp},i} = \sqrt{\sum_{k=1}^{N_i} a_{ik}^2I_{ik}(I_{ik}+1)}.
\end{equation}
For PTZ$^{\bullet+}$, the experimental hyperfine field is 0.96 mT, while the calculations suggest an effective field of 0.64 mT; for PDI$^{\bullet-}$, the fields are 0.27 mT and 0.34 mT respectively. The results for the PTZ$^{\bullet+}$ radical are particularly poor. This may in part be due to the fact the EPR-II basis set normally used to calculate hyperfine constants cannot be employed for this sulphur-containing radical, as it is only parametrised for Period II elements.\cite{Barone96} For this radical we used the larger but hyperfine-unoptimised cc-PV5Z orbital basis set instead.\cite{Dunning89,Woon93} 

The DFT calculations do at least provide the signs of the hyperfine coupling constants, which are not available from the experimental EPR\cite{Chiu70} or ENDOR\cite{Tauber06} data. In the simulations reported below, we shall use the magnitudes of the hyperfine coupling constants from the experiments and the signs from the DFT calculations. This combination will enable us to obtain a good fit to the magnetic field effects (MFEs) in the triplet and radical pair yields of the charge recombination reaction observed by Weiss {\em et al.}\cite{Weiss04} With the raw DFT hyperfine coupling constants it is simply not possible to obtain such a good fit.

\begin{figure}[t]
\centering
\resizebox{0.7\columnwidth}{!} {\includegraphics{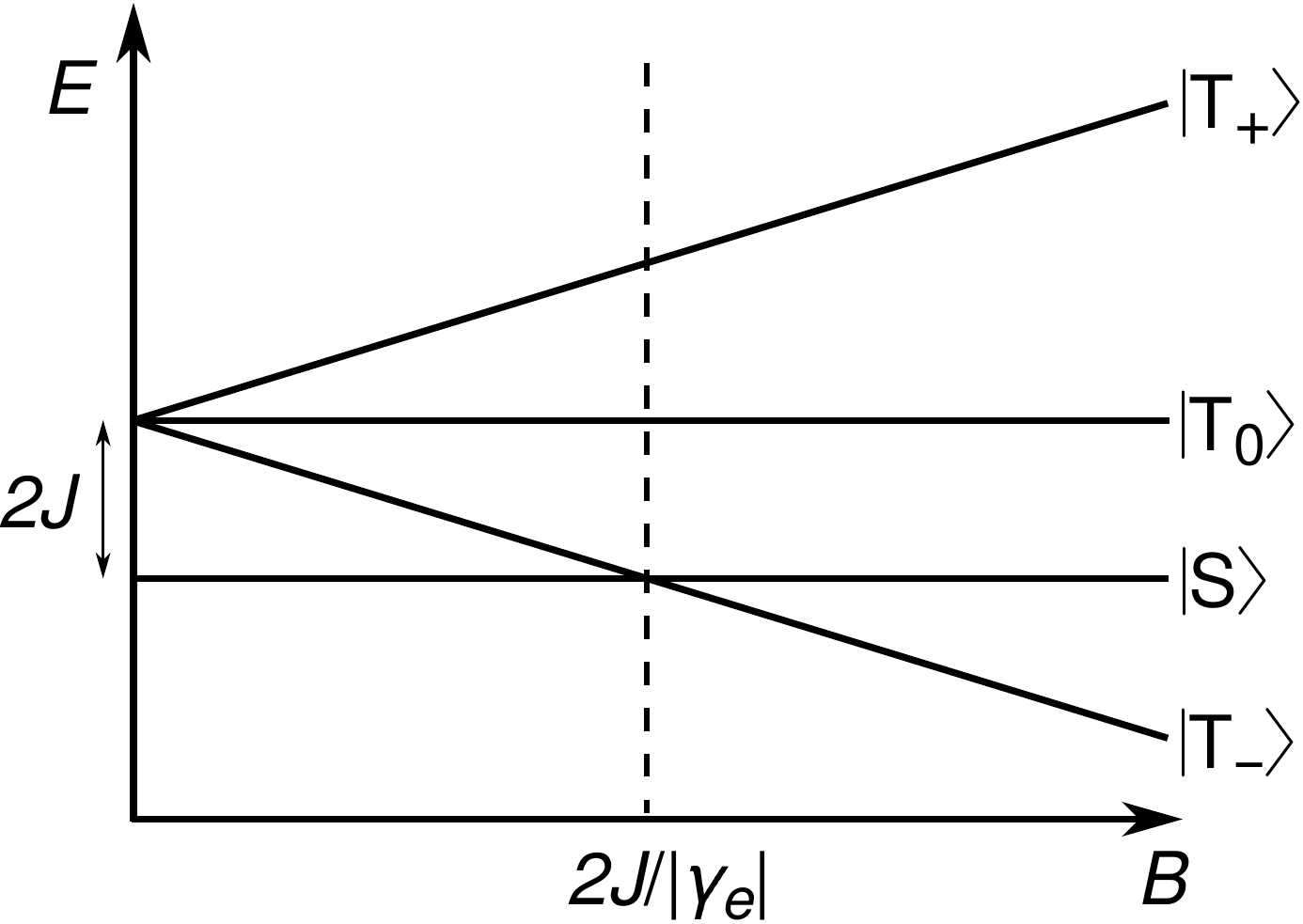}}
\caption{A schematic representation of the energy levels of the spin states of the radical pair as a function of the strength of the applied magnetic field, $B$, ignoring the effect of hyperfine interactions.}
\label{Energies}
\end{figure}

\begin{table}[b]
\centering
\begin{tabular} { c  c  c  c  c }
\quad $n$\quad &\quad 2\quad &\quad 3\quad &\quad 4\quad &\quad 5\quad \\
\hline
\hline
\quad $2J/|\gamma_{\rm e}|$ in mT\quad &\quad 170\quad &\quad 31\quad &\quad 6.4\quad &\quad 1.5\quad \\
\quad $\tau$ in ns\quad &\quad 21\quad &\quad 330\quad &\quad 217\quad &\quad 121\quad \\
\hline
\hline
\end{tabular}
\caption{Exchange coupling constants and zero-field charge recombination lifetimes of PTZ$^{\bullet+}$--Ph$_n$--PDI$^{\bullet-}$ molecular wires, both taken from Ref.~\onlinecite{Weiss04}.}
\label{Lifetimes}
\end{table}

The exchange coupling constants $J$ of the PTZ$^{\bullet+}$--Ph$_n$--PDI$^{\bullet-}$ molecular wires can be extracted from the MFE measurements that we have just mentioned. In their experiments, Weiss {\em et al.} found that the triplet yields of the shorter wires ($n=2,3$) went through a maximum, and that the radical pair yields of the longer wires ($n=4,5$) went through a minimum, as the strength of the applied magnetic field was increased.\cite{Weiss04} Both of these observations can be rationalised in terms of the relative energy levels of the singlet and triplet electronic states, shown schematically in Fig.~\ref{Energies}. When there is no applied field, or if the applied field is very large, the singlet state is separated in energy from the triplet states, limiting singlet-triplet interconversion. However, on resonance, where $B=2J/|\gamma_{\rm e}|$, the singlet state in which the radical pair is formed becomes degenerate with the $\ket{\rm T_-}$ state. This results in more efficient intersystem crossing and a maximum in the triplet yield. Since $k_{\rm T}> k_{\rm S}$ (see below), it also results in a minimum in the \lq\lq radical pair yield" (i.e., the radical pair survival probability 50 ns after the initial photoexcitation laser pulse). Therefore, by determining the strength of the applied field at which the triplet (radical pair) yield is largest (smallest), it is possible to infer the magnitude of $2J$. The values of $2J$ for each of the wires were determined in this way in Ref.~\onlinecite{Weiss04}, and are given again here in Table \ref{Lifetimes}. 

(Note that, in drawing Fig.~3, we have assumed that the exchange coupling is antiferromagnetic ($J>0$). This is consistent with the results of time-resolved EPR measurements of electron spin polarization in these molecular wires.\cite{Dance06} If the exchange coupling were ferromagnetic ($J<0$), the resonance condition would be $B=-2J/|\gamma_{\rm e}|$, and the $\ket{\rm T_-}$ state would be replaced by the $\ket{\rm T_+}$ state. However, the MFEs would be the same: one would still see a maximum in the triplet yield and a minimum in the radical pair yield on resonance.)

\subsection{Recombination operator}

\begin{figure}[t]
\centering
\resizebox{0.8\columnwidth}{!} {\includegraphics{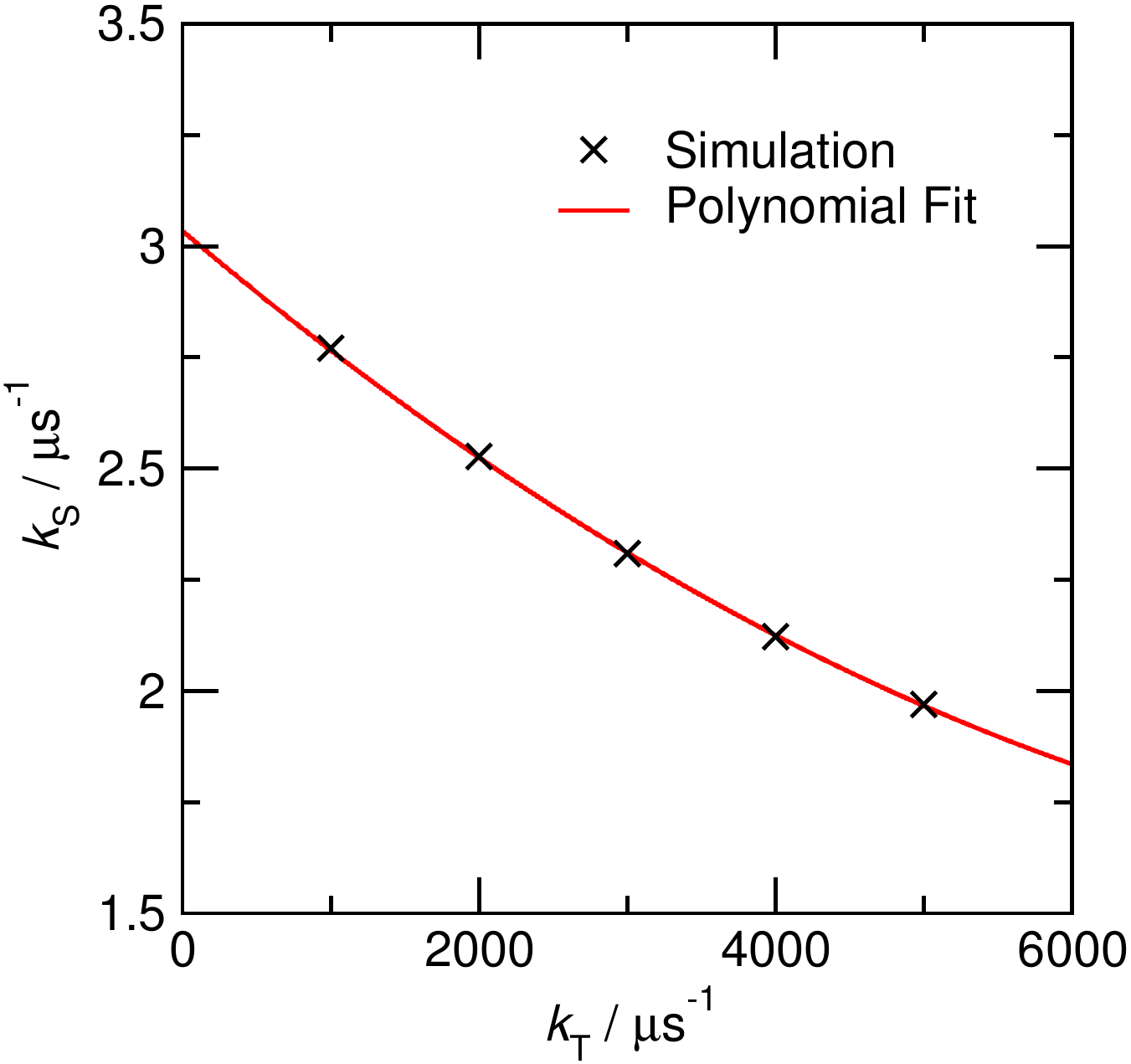}}
\caption{Five pairs of values of $(k_{\rm S},k_{\rm T})$ which reproduce the overall charge recombination lifetime of PTZ$^{\bullet+}$--Ph$_3$--PDI$^{\bullet-}$ in the absence of an applied magnetic field are shown in black. The polynomial fit to these points is shown in red, and is defined by $k_{\rm S} = ak_{\rm T}^2 + bk_{\rm T} + c$. The coefficients $a,$ $b,$ and $c$ for all four molecular wires ($n=2-5$) are given in Table \ref{Parameters}.}
\label{ParameterSpace}
\end{figure}

The singlet and triplet charge recombination rate constants $k_{\rm S}$ and $k_{\rm T}$ have not been measured directly for PTZ$^{\bullet+}$--Ph$_n$--PDI$^{\bullet-}$ molecular wires, but there are some experimental observations which provide constraints on them. Firstly, as we have already mentioned above, a minimum in the radical pair yield at the resonance point, rather than a maximum, implies that $k_{\rm T}>k_{\rm S}$. Secondly, the charge separated radical pair lifetime of each molecular wire has been measured in the absence of a magnetic field by monitoring the decay of the 720 nm absorption band of PDI$^{\bullet-}$.\cite{Weiss04} These lifetimes are also given in Table \ref{Lifetimes}. 

For a given molecular wire, the charge separated radical pair lifetime provides one constraint on the two unknowns $k_{\rm S}$ and $k_{\rm T}$. This is illustrated in Fig.~4, which shows the line in the $k_{\rm S}$, $k_{\rm T}$ plane for the PTZ$^{\bullet+}$--Ph$_3$--PDI$^{\bullet-}$ molecular wire along which our spin dynamics calculations reproduce the experimental radical pair lifetime of $\tau=330$ ns in the absence of an applied magnetic field. This line is well fit by writing $k_{\rm S}$ as a quadratic function of $k_{\rm T}$, and we have found the same to be true for all of the other molecular wires. The resulting quadratic fits are summarised in Table~IV. Since these fits furnish $k_{\rm S}$ for a given $k_{\rm T}$, we are left with a single free parameter to vary to reproduce the MFEs observed by Weiss {\em et al.}\cite{Weiss04} for each wire.
\begin{table}[b]
\centering
\begin{tabular} { c  c  c  c }
\quad $n$\quad &\quad $a / \mu$s\quad &\quad $b$\quad &\quad $c / \mu$s$^{-1}$\quad \\
\hline
\hline
\quad 2 &\quad $8.540 \times 10^{-11}$ &\quad $-9.375 \times 10^{-6}$ &\quad 47.620 \\
\quad 3 &\quad $1.372 \times 10^{-8}$ &\quad $-2.821 \times 10^{-4}$ &\quad 3.034  \\
\quad 4 &\quad $2.116 \times 10^{-6}$ &\quad $-6.983 \times 10^{-3}$ &\quad 4.634 \\
\quad 5 &\quad $1.194 \times 10^{-3}$ &\quad $-0.215$ &\quad 9.923 \\
\hline
\hline
\end{tabular}
\caption{The coefficients of the polynomial $k_{\rm S} = ak_{\rm T}^2 + bk_{\rm T} + c$ which defines the $(k_{\rm S},k_{\rm T})$ parameter space consistent the experimental radical pair lifetime of each molecular wire in the absence of an applied magnetic field. }
\label{Parameters}
\end{table}

\section{Results}

\subsection{Shorter wires}
\label{Short}

\begin{figure}[t]
\centering
\resizebox{0.9\columnwidth}{!} {\includegraphics{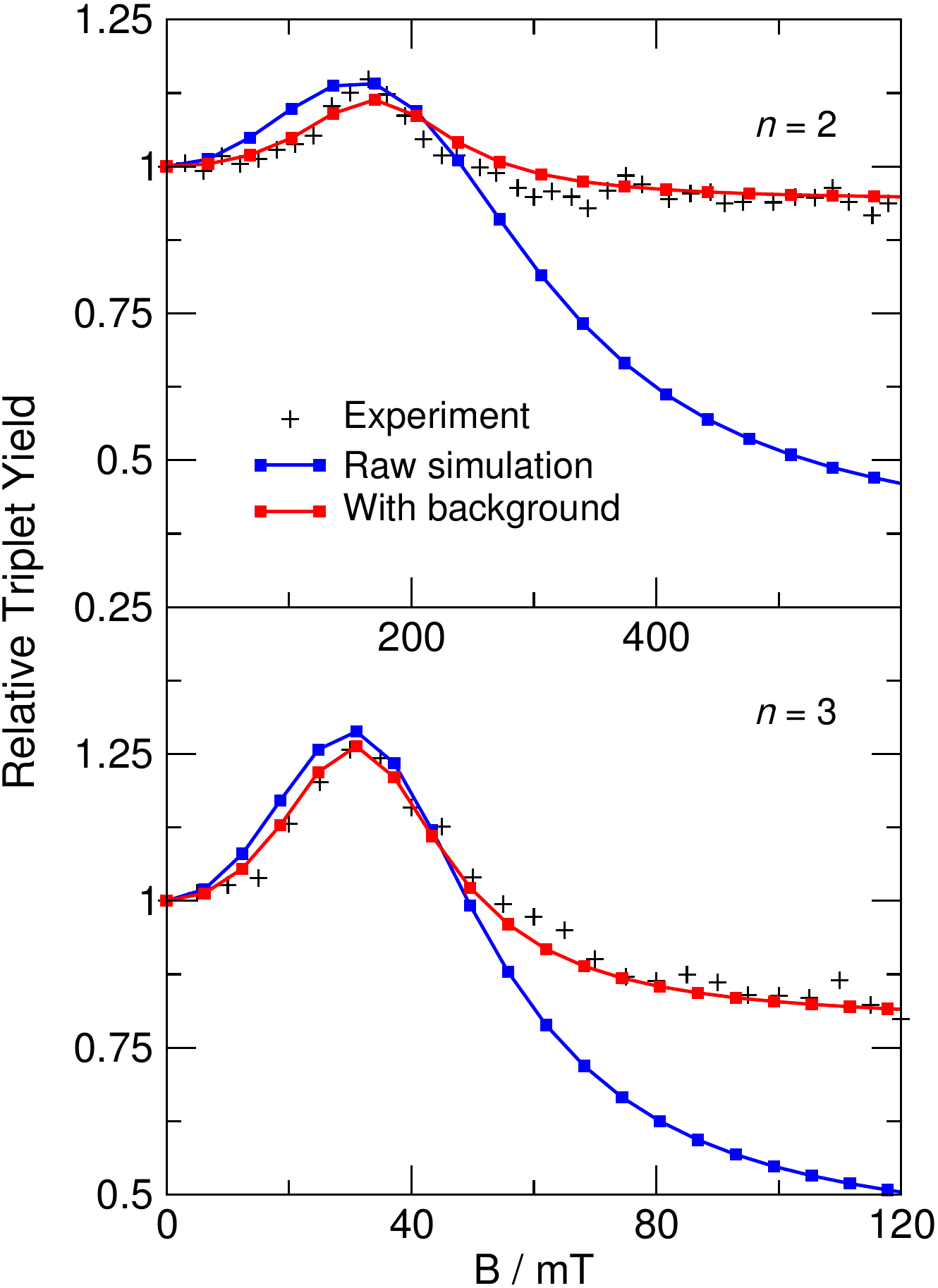}}
\caption{The triplet yield of PTZ$^{\bullet+}$--Ph$_n$--PDI$^{\bullet-}$ as a function of the strength of the applied magnetic field, relative to the triplet yield in the absence of a field, for $n = 2$ above and $n = 3$ below. The blue curves show the results of our raw spin dynamics simulations, and the red curves the results of simulations with a field-independent background contribution to the triplet yield. The experimental data is taken from Ref.~\onlinecite{Weiss04}.}
\label{TY}
\end{figure}

For the wires with $n=2$ and 3, Weiss {\em et al.}\cite{Weiss04}  measured the triplet yield of the radical pair recombination reaction as a function of the strength of the applied magnetic field, and reported this as the relative triplet yield $\Phi_{\rm T}(B)/\Phi_{\rm T}(0)$. We have simulated these experiments using the parameters in the spin Hamiltonian and recombination operator defined above, using $k_{\rm T}$ as an adjustable parameter to fit the experimental data and specifying $k_{\rm S}$ in terms of $k_{\rm T}$ in accordance with Table~IV. The resulting least squares fits to the experimental $\Phi_{\rm T}(B)/\Phi_{\rm T}(0)$ curves are shown in Fig.~5 in blue. 

While the positions and heights of the resonance peaks in the relative triplet yields of both wires are captured well by these calculations, it is clear that there is very poor agreement between theory and experiment in the high field region. This discrepancy is interesting. We believe it suggests that the photochemical scheme in Fig.~1 is incomplete for these molecular wires, and in particular that it is missing a magnetic field-independent background contribution to the triplet yield (the yield of $^3$D-B-A$^*$ in Fig.~1). 

In the absence of such a background contribution, and assuming as we have done in our calculations that the $\Delta g$ mechanism of intersystem crossing can be discounted, one would expect the high field limit of the triplet yield to be approximately one third of the zero field value. This follows from Fig.~3. When $B=0$, all three components of the triplet state have the same energy gap to the singlet state, and are therefore equally energetically accessible to hyperfine-mediated intersystem crossing. But for sufficiently large $B$, the $\left|{\rm T}_+\right>$ and $\left|{\rm T}_-\right>$ triplet components become energetically inaccessible, leaving only a third as many pathways for singlet to triplet conversion. This simple picture is consistent with the raw spin dynamics results in Fig.~5, which are seen to be tending towards a relative triplet yield of around a third in the high field limit. However, it is manifestly inconsistent with the experimentally measured relative triplet yields of Weiss {\em et al.}\cite{Weiss04} 

As we have already suggested, the discrepancy can be resolved by assuming that the experiment is detecting an additional contribution to the $^3$D-B-A$^*$ yield that is produced by some magnetic field-independent process outside of the mechanism outlined in Fig.~\ref{RPR}. This would in effect add a  `background' contribution to the triplet yields calculated using the radical pair model. To allow for such a contribution, we have recalculated the relative triplet yield (RTY) as
\begin{equation}
{\rm RTY}(B) = \frac{\Phi_{\rm T}(B) + x}{\Phi_{\rm T}(0) + x},
\label{background}
\end{equation}
where $\Phi_{\rm T}(B)$ is the simulated triplet yield at magnetic field strength $B$, and $x$ is the background contribution, defined as
\begin{equation}
x = \frac{\lambda\,\Phi_{\rm T}(0) - \Phi_{\rm T}(\infty)}{1 - \lambda}, 
\label{x}
\end{equation}
where $\lambda$ is the experimental high field limit of the relative triplet yield. 

New least squares fits to the experimental relative triplet yields were found by using Eq.~\eqref{background} to re-optimise the triplet recombination rate constants $k_{\rm T}$ for the $n=2$ and 3 wires in the spin dynamics simulations. The resulting least squares fits are plotted in red in Fig.~\ref{TY}. With the background corrections included, the simulations agree quantitatively with the experiments, allowing us to extract optimum values of $k_{\rm T}$ (and therefore also $k_{\rm S}$ -- see Table~IV) for both molecular wires. These are listed in Table \ref{T_Rates}. The empirical parameters used in the background corrections were $\lambda = 0.946$ for $n = 2$ and $\lambda = 0.814$ for $n = 3$, giving $x=0.0416$ and $x=0.525$, respectively. A strong \emph{a posteriori} justification for including the background corrections will be presented in Sec.~V.C, and several possible magnetic field-independent mechanisms by which the $^3$D-B-A$^*$  state could have be produced in the experiments will be discussed in Sec.~V.D.

\begin{table} [b]
\centering
\begin{tabular} { c  c  c }
\quad $n$\quad &\quad $k_{\rm T}$ / $\mu$s$^{-1}$\quad &\quad $k_{\rm S}$\quad / $\mu$s$^{-1}$ \\
\hline
\hline
\quad2\quad & 27500 & 47.4 \\
\quad3\quad & 3800 & 2.16 \\
\quad4\quad & 350 & 2.45 \\
\quad5\quad & 60.0 & 2.89 \\
\hline
\hline
\end{tabular}
\caption{The singlet and triplet charge recombination rate constants of the PTZ$^{\bullet+}$--Ph$_n$--PDI$^{\bullet-}$ molecular wires obtained from the present calculations.}
\label{T_Rates}
\end{table}

\subsection{Longer wires}
\label{Long}

\begin{figure}[t]
\centering
\resizebox{0.9\columnwidth}{!} {\includegraphics{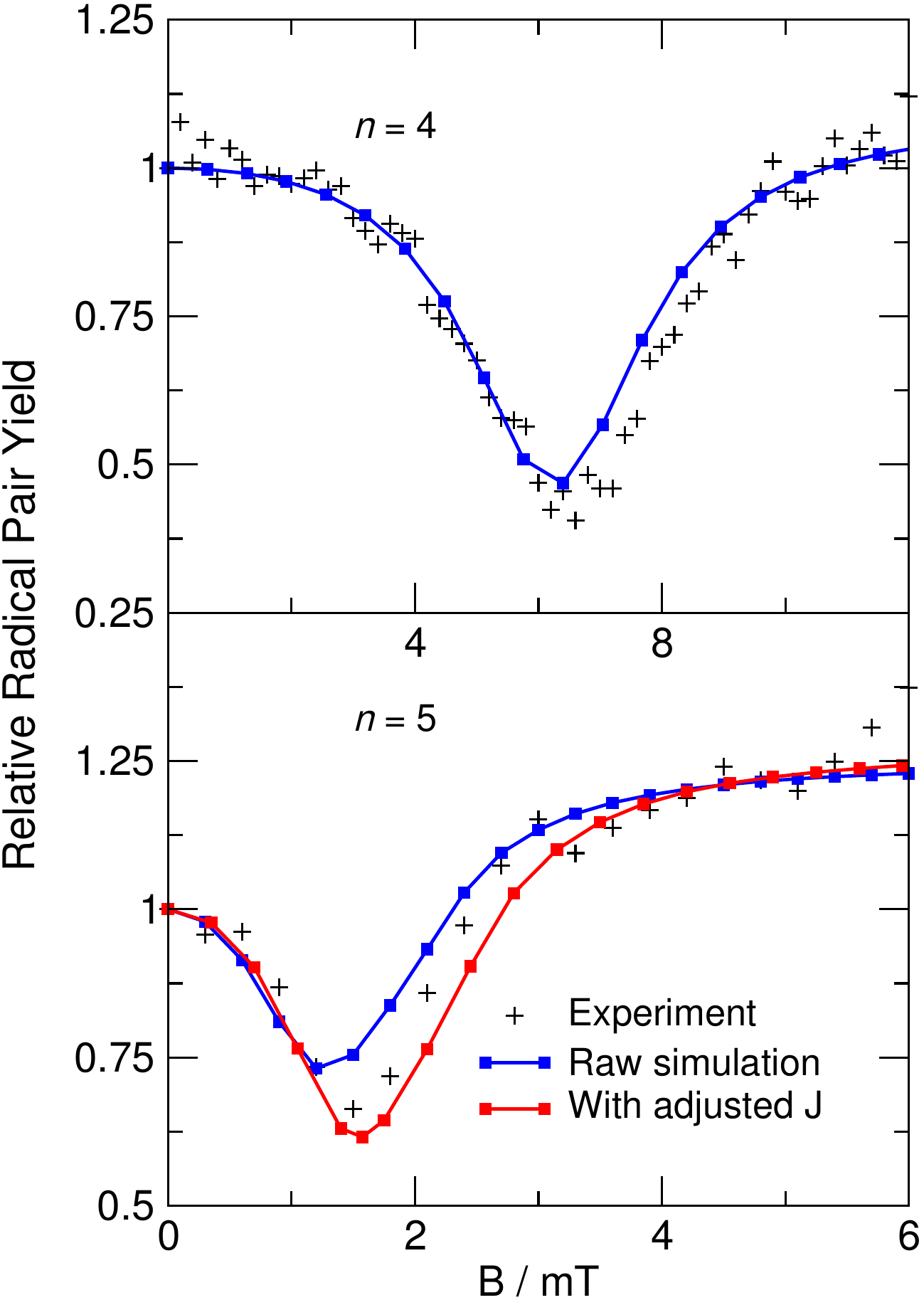}}
\caption{The radical pair yield of PTZ$^{\bullet+}$--Ph$_n$--PDI$^{\bullet-}$ as a function of the strength of the applied magnetic field, relative to the radical pair yield in the absence of a field, for $n = 4$ above and $n = 5$ below. The blue curves were obtained using the exchange coupling constants given in Ref.~\onlinecite{Weiss04}; the red curve in the lower panel was obtained with $2J / |\gamma_{\rm e}| = 1.75$ mT rather than 1.50 mT. The experimental data is taken from Ref.~\onlinecite{Weiss04}.}
\label{RP}
\end{figure}

For the wires with $n = 4$ and 5, Weiss {\em et al.}\cite{Weiss04} measured the radical pair yield 50 ns after the initial photoexcitation laser pulse as a function of the strength of the applied magnetic field, and again reported this as the relative radical pair yield $\Phi_{\rm RP}(B) / \Phi_{\rm RP}(0)$. In simulating these results, we found a least squares fit to the experimental data at $t = 55$ ns rather than 50 ns, most likely because of the finite (7 ns) experimental instrument response time.\cite{Weiss04} Our results are plotted in Fig.~\ref{RP} in blue. For $n = 4$, excellent quantitative agreement is observed between simulation and experiment without the need for any further correction. However, the results for $n = 5$ are not as good, with the simulated minimum in the radical pair yield at a different magnetic field strength than that observed experimentally.

When $n = 5$, the exchange coupling is comparable to the sum of the effective hyperfine fields in the two radicals (0.96 mT in PTZ$^{\bullet+}$ and 0.27 mT in PDI$^{\bullet-}$). As a result, intersystem crossing to the $\ket{\rm T_+}$ state cannot entirely be neglected at the point where the $\ket{\rm T_-}$ state comes into resonance with the $\ket{\rm S}$ state, as it can for the shorter wires. As the magnetic field strength increases towards $2J/|\gamma_{\rm e}|$, the energy gap between the $\ket{\rm S}$ and $\ket{\rm T_+}$ states increases, reducing the rate of transition between these two states. At the same time, the rate of crossing from $\ket{\rm S}$ to $\ket{\rm T_-}$ increases, becoming most efficient when $B= 2J/|\gamma_{\rm e}|$. Therefore, the total intersystem crossing is most efficient, and a minimum in the radical pair yield is observed, at a field strength somewhat below $2J/|\gamma_{\rm e}|$. 

Because of this, one cannot simply read off the magnitude of the exchange coupling constant $J$ from the magnetic field strength at the minimum in the radical pair yield for the $n=5$ wire, as was done by Weiss {\em et al.}.\cite{Weiss04} The comparable magnitudes of the exchange and hyperfine interactions require $J$ to be extracted from a spin dynamics calculation. To do this, we varied $J$ in our simulations until the position of the minimum in the computed radical pair yield matched the experimental data.\cite{Footnote} Using the resulting optimised value of $J$ ($2J/|\gamma_{\rm e}|=1.75$ mT), a new lifetime-constrained $(k_{\rm S},k_{\rm T})$ parameter space was constructed (with $a=0.652\times 10^{-3}$ $\mu$s, $b=-0.147$, and $c=9.377$ $\mu$s$^{-1}$), and from this a least squares fit to the experimental radical pair yield was obtained by varying $k_{\rm T}$. This least squares fit is shown as the red curve in the lower panel of Fig.~\ref{RP}. Our final optimised values of the singlet and triplet charge recombination rate constants $k_{\rm S}$ and $k_{\rm T}$ for both of the longer wires are given along with those of the shorter wires in Table~\ref{T_Rates}.

\section{Discussion}

\subsection{Role of the $\Delta g$ mechanism}

The simulations we have just presented will enable us to investigate several interesting questions about the physics of the charge recombination along these PTZ$^{\bullet+}$--Ph$_n$--PDI$^{\bullet-}$ molecular wires. But to begin with, let us return to the issue of the $g$ factors of the electrons in the radical pair discussed in Sec.~III.A. 

\begin{figure}[t]
\centering
\resizebox{0.9\columnwidth}{!} {\includegraphics{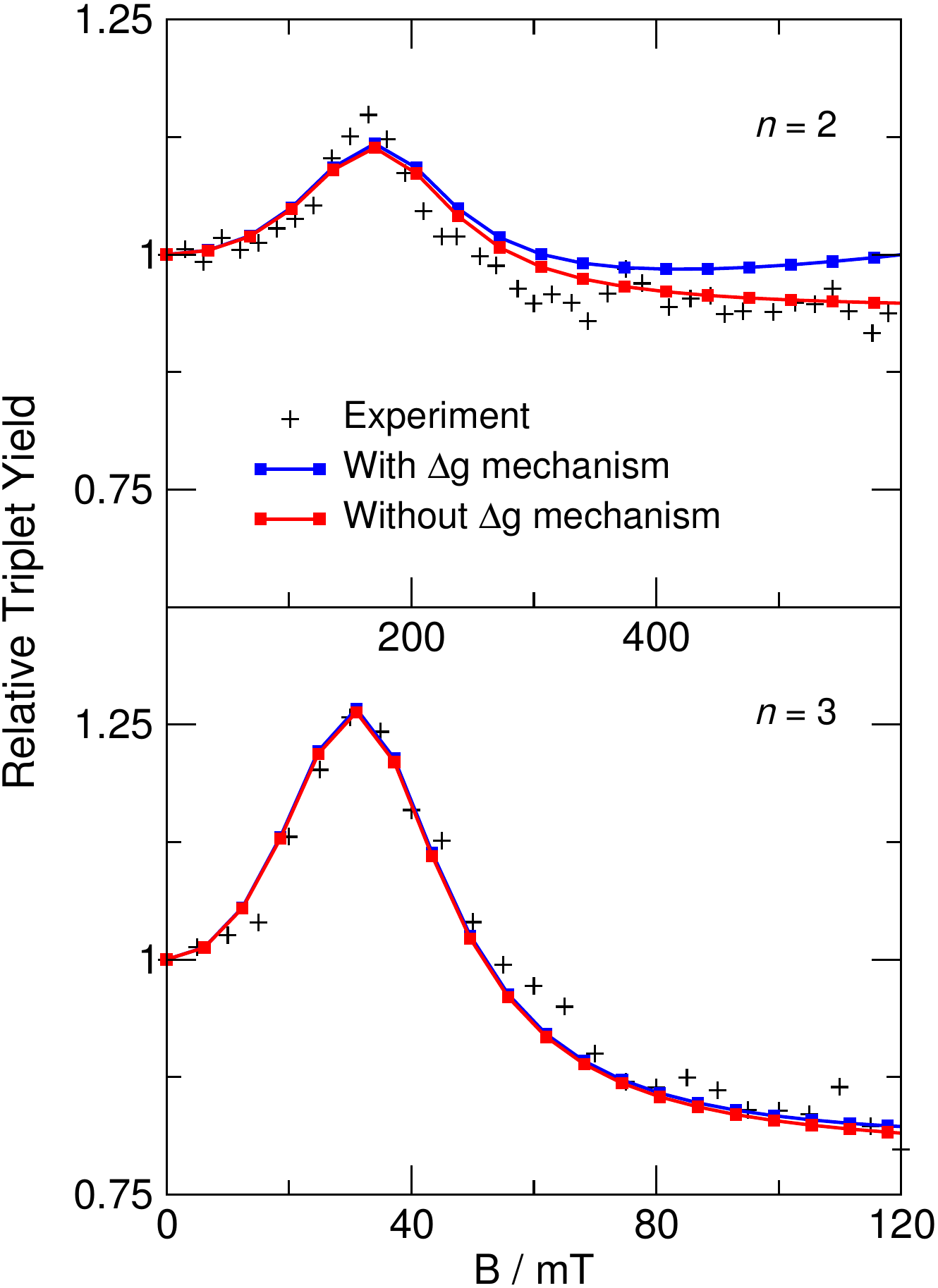}}
\caption{Relative triplet yields as in Fig.~5, with the magnetic field-independent background corrections included, comparing the results of simulations with $g_1=g_2=g_{\rm e}=2.0023$ and simulations with $g_1=2.0053$ and $g_2=2.0028$.}
\end{figure}

In all of the calculations we have presented so far, we have simply set $g_1=g_2=g_{\rm e}$. In reality, the $g$ factors of the two unpaired electrons will be slightly different, although perhaps not quite so different as they are in the isolated PTZ$^{\bullet +}$ and PDI$^{\bullet -}$ radicals. The difference $\Delta g$ between the $g$ factors introduces a new term $\Delta\hat{H}=\Delta g\mu_{\rm B}B(\hat{S}_{1z}-\hat{S}_{2z})/\hbar$ into the spin Hamiltonian, which causes transitions between the $\ket{{\rm S}}$ and $\ket{{\rm T}_0}$ states. According to Fermi's Golden Rule, the rate of intersystem crossing induced by this mechanism will be proportional to $\vert\bra{{\rm T}_0}\Delta\hat{H}\ket{{\rm S}}\vert^2$, and therefore to $B^2$. Even if $\Delta g$ is very small, this will clearly have some effect on the magnetic field dependence of the triplet and radical pair yields at sufficiently high magnetic fields.

In order to quantify this effect, we have repeated our calculations with the isolated radical $g$ factors,\cite{Chiu70,Tauber06} $g_1=2.0053$ for PTZ$^{\bullet+}$ and $g_2=2.0028$ for PDI$^{\bullet-}$.  The radical pair yields of the longer wires in Fig.~6, and the triplet yield of the $n=3$ wire in Fig.~5, were found to be unchanged to graphical accuracy. However, the triplet yield of the $n=2$ wire was found to increase slightly as a result of the $\Delta g$ mechanism beyond $B=200$ mT. This increase is shown in Fig.~7, which also shows the negligible effect of the mechanism on the triplet yield of the $n=3$ wire below $B=120$ mT. 

The $n=2$ results in Fig.~7 seem to suggest that the $g$ factors of the electrons in the isolated PTZ$^{\bullet+}$ and PDI$^{\bullet-}$ radicals give a $\Delta g$ that is too large to be compatible with the PTZ$^{\bullet+}$--Ph$_2$--PDI$^{\bullet-}$ experiments of Weiss {\em et al.}\cite{Weiss05} When the $\ket{{\rm S}}\to\ket{{\rm T}_0}$ intersystem crossing is included with this $\Delta g$ value, the simulated triplet yield starts to increase beyond $B=400$ mT, whereas the experimental triplet yield has reached a plateau (or is perhaps even decreasing slightly) at this field strength. Presumably, this is because the -Ph$_2$-PDI$^{\bullet-}$ substituent on the PTZ$^{\bullet+}$ radical alters its $g$ factor, and the PTZ$^{\bullet+}$-Ph$_2$ substituent on the PDI$^{\bullet-}$ radical alters its $g$ factor, bringing the $\Delta g$ value in the radical pair closer to zero (the red curve in Fig.~7).

\subsection{Resonance peak widths}

The first question we shall use our simulations to answer is why the resonance peaks in the triplet and radical pair yield MFEs in Figs.~5 and~6 are so broad. Previous experimental studies of similar D$^{\bullet+}$-B-A$^{\bullet-}$ molecular wires have noted that the widths of the MFE peaks are often far larger than the hyperfine interactions in the radical pair,\cite{Weiss03,Weiss05} and that is also the case here. The sum of the hyperfine fields in the PTZ$^{\bullet+}$ and PDI$^{\bullet-}$ radicals is just $0.96+0.27=1.23$ mT, whereas the widths of peaks in the triplet yield MFE in Fig.~5 are $\sim 100$ mT for $n=2$ and $\sim 25$ mT for $n = 3$. It is therefore implausible that hyperfine interactions alone could be responsible for the observed peak widths.

\begin{figure}[t]
\centering
\resizebox{0.8\columnwidth}{!} {\includegraphics{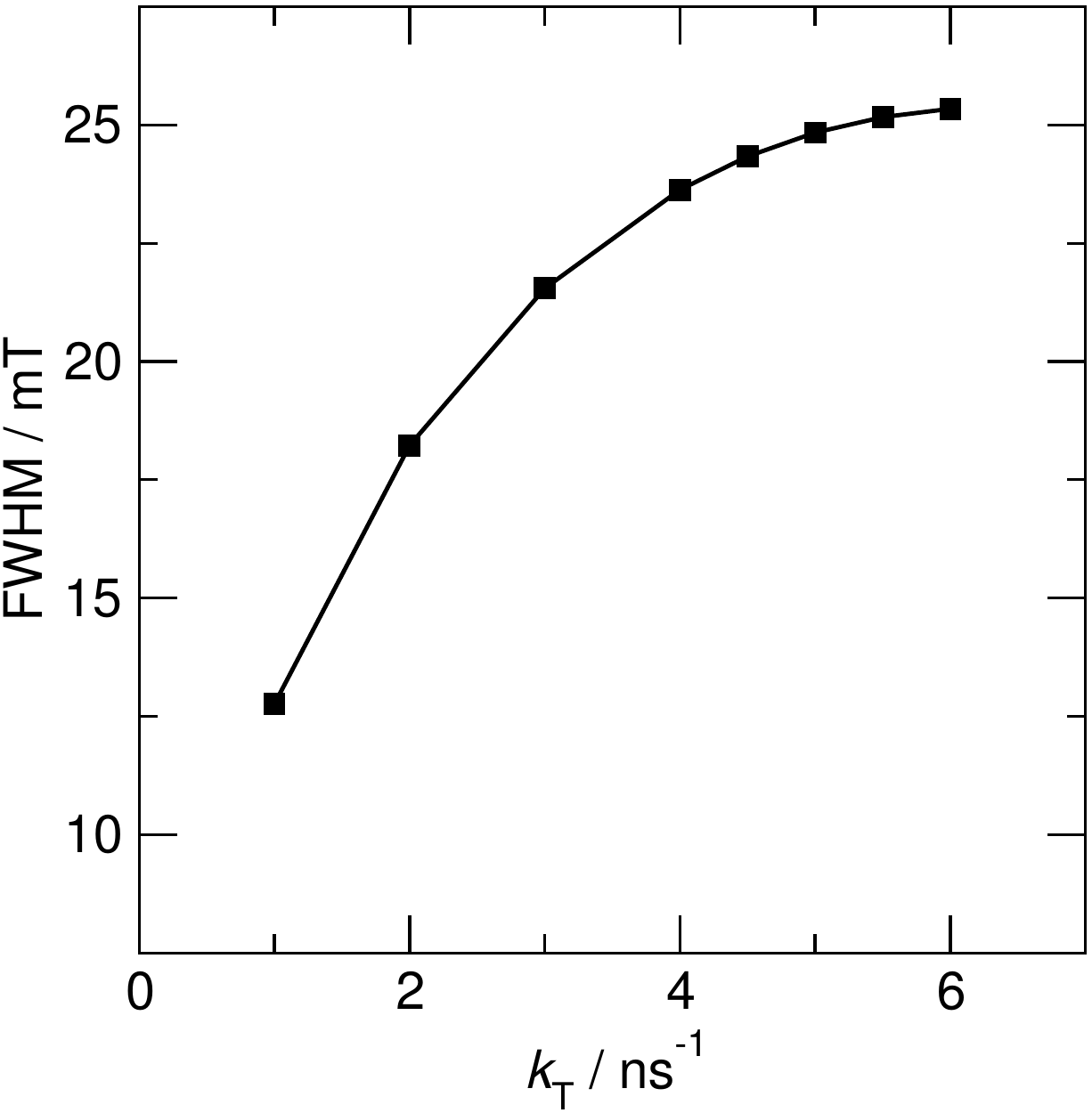}}
\caption{The full width at half maximum of the simulated peak in the triplet yield of PTZ$^{\bullet+}$--Ph$_3$--PDI$^{\bullet-}$ as a function of the triplet recombination rate constant, $k_{\rm T}$. For each point, $k_{\rm S}$ is chosen to give the correct zero-field lifetime of the radical pair in accordance with Fig.~4.}
\label{FWHM}
\end{figure}

Our calculations suggest that the resonance widths are dominated by the lifetime broadening of the triplet state of the charge-separated radical 
pair.\cite{Hoff81} Indeed dividing our triplet charge recombination rate constants for the wires with $n=2$ and $3$ by $|\gamma_{\rm e}|=176.1\ \mu{\rm s}^{-1}{\rm mT}^{-1}$ gives magnetic field strengths of 156 mT and 22 mT, respectively, which are of the same orders of magnitude as the observed resonance widths in Fig.~5. The short lifetime of the triplet states of the radical pair leads to a broadening of their energy levels, giving a non-zero density of triplet states at the energy of the singlet state over a wide range of magnetic field strengths around the resonance at $B=2J/|\gamma_{\rm e}|$. The singlet state has a much longer lifetime (see Table~V), and so we do not expect lifetime broadening will have such a significant effect on its density of states. 

Figure~\ref{FWHM} supports this explanation: for the $n = 3$ wire, the full width at half maximum (FWHM) of the peak in the simulated triplet yield increases monotonically with increasing $k_{\rm T}$. Since the triplet charge recombination rate constant decreases as $n$ increases, it is likely that when $n = 4$ hyperfine interactions will make some contribution to the width of the resonance (along with lifetime broadening), and when $n = 5$ hyperfine interactions will certainly contribute to the resonance width.

\begin{figure}[t]
\centering
\resizebox{0.8\columnwidth}{!} {\includegraphics{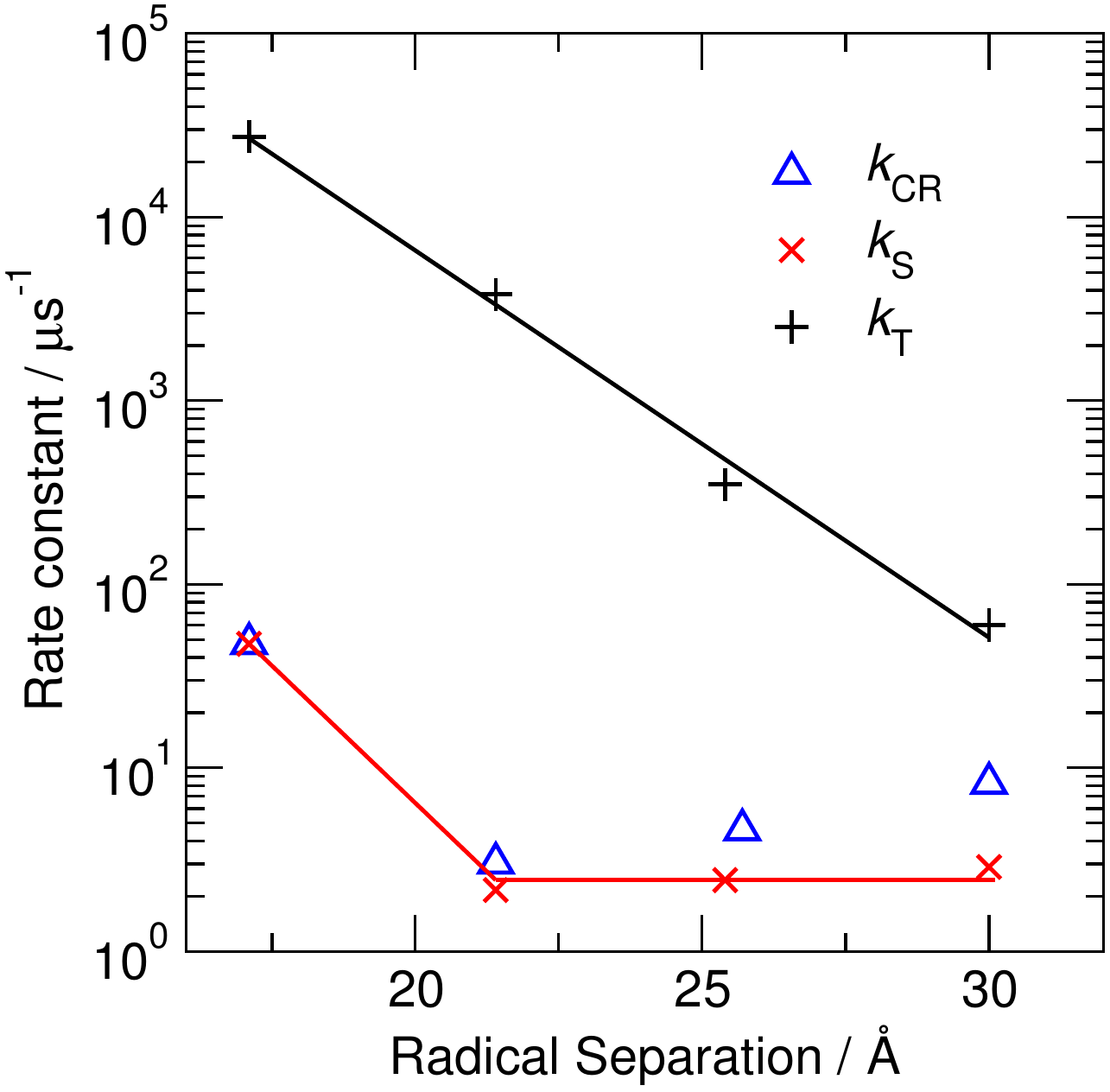}}
\caption{The singlet and triplet charge recombination rate constants $k_{\rm S}$ and $k_{\rm T}$ of PTZ$^{\bullet+}$--Ph$_n$--PDI$^{\bullet-}$ for $n = 2 - 5$ extracted from our simulations, plotted as a function of the radical pair separation in these wires. $k_{\rm T}$ follows a single exponential with decay constant $\beta_{\rm T} = 0.48\ {\rm \AA}^{-1}$. The overall (zero magnetic field) charge recombination rate constants $k_{\rm CR}$ are the reciprocals of the radical pair lifetimes in Table \ref{Lifetimes}. These are taken from Ref.~\onlinecite{Weiss04}.}
\label{Rates}
\end{figure}

\subsection{Singlet and triplet recombination mechanisms}

The second question we can address is the mechanism of the charge recombination along the singlet and triplet pathways. Figure~\ref{Rates} shows how $k_{\rm S}$ and $k_{\rm T}$ vary as a function of the distance between the radicals in the pair, with the rates plotted on a logarithmic scale. The triplet recombination rate decreases exponentially with the radical separation, with a decay constant $\beta_{\rm T} = 0.48$ ${\rm \AA}^{-1}$. This is very similar to that observed by Weiss {\em et al.}\cite{Weiss04} for the initial charge separation, $\beta_{\rm CS}=0.46$  ${\rm \AA}^{-1}$. Since the exponential dependence is characteristic of the superexchange mechanism we conclude that the triplet charge recombination occurs by superexchange in all four molecular wires. 

Note in passing that the triplet recombination rate constants in Fig.~\ref{Rates} provide an {\em a posteriori} justification for the background corrections to the triplet yields of the $n=2$ and 3 wires that we introduced in Sec.~IV.A: the resulting values of $k_{\rm T}$ are entirely consistent with those for the $n=4$ and 5 wires, which were obtained independently from the experimental radical pair yields in Fig.~6 (which do not involve any background correction). The quality of the single exponential fit to all four $k_{\rm T}$ data points in Fig.~\ref{Rates} ($R^2=0.997$) certainly supports this.

The singlet recombination rate constants in Fig.~\ref{Rates} are very similar for the $n = 3 - 5$ wires, but $k_{\rm S}$ is significantly larger for $n=2$. This suggests a change in the mechanism of the singlet charge recombination pathway as the bridge length increases, with the superexchange mechanism making a significant contribution for $n=2$ and the incoherent hopping mechanism dominating thereafter. 

This change in mechanism can be understood in terms of Marcus theory. The direct recombination of the singlet radical pair lies deep in the Marcus inverted region,\cite{Marcus65} which disfavours the superexchange mechanism.\cite{Weiss04} For wires with short bridges, the large electronic coupling between the electron donor and acceptor can compensate for this, such that when $n = 2$ superexchange still dominates. However, as the bridge length increases and the electronic coupling decreases, the superexchange mechanism becomes slow compared to the incoherent hopping mechanism. Direct recombination of the triplet radical pair is not as deep in the inverted region, because the triplet product ($^3$D-B-A$^*$) is higher in energy than the singlet product ($^1$D-B-A).\cite{Weiss04} As a result, the superexchange mechanism is far more favourable for the triplet radical pairs, and dominates for all bridge lengths.

It is important to note that these insights could not have been obtained from the overall (zero field) experimental recombination rate constants $k_{\rm CR}$ alone. These depend not only on $k_{\rm S}$ and $k_{\rm T}$, but also on the rate of intersystem crossing between the spin states of the radical pair. This can clearly be seen by comparing $k_{\rm S}$ and $k_{\rm CR}$ for the $n = 2$ and $n = 5$ wires in Fig.~\ref{Rates}. When $n = 2$, the exchange coupling that sets the energy gap between the singlet and triplet states is large, so intersystem crossing from the singlet state to the triplet state in zero field is slow, and $k_{\rm CR} \approx k_{\rm S}$. However, when $n = 5$, the exchange coupling is much smaller and intersystem crossing is much more efficient, so recombination of the triplet radical pair contributes significantly to the overall recombination rate and $k_{\rm CR} > k_{\rm S}$. 

It is also clear from this argument that $k_{\rm CR}$ will depend on the strength of the applied magnetic field, whereas $k_{\rm S}$ and $k_{\rm T}$ do not. In fact, this allows us to make a prediction. If the experimental measurements of $k_{\rm CR}$ in Ref.~\onlinecite{Weiss04} were to be repeated in the presence of an applied magnetic field, we would expect them to satisfy $k_{\rm T}>k_{\rm CR}(B)\ge k_{\rm S}$ for all field strengths $B$, and to approach the value we have obtained for $k_{\rm T}$ most closely for each wire at the resonant field strength $B=2J/|\gamma_{\rm e}|$. However, $k_{\rm CR}(B)$ will never actually reach $k_{\rm T}$, because the triplet yield of the charge recombination reaction never reaches 1. This clearly precludes a direct experimental measurement of $k_{\rm S}$ and $k_{\rm T}$ by this method. Insofar as we can see, these rate constants can only be determined by fitting the magnetic field dependence of the experimental results to the results of quantum spin dynamics simulations, as we have done in this paper.

\subsection{Background contribution to the triplet yield}

The last question raised by our results is perhaps the most interesting, because we have yet to find a satisfactory answer to it. What is the physical origin of the magnetic field-independent background contribution to the triplet yield of the charge recombination reaction that we introduced in Sec.~IV.A (and have since justified in Sec.~V.C)? Here we will discuss three possible explanations for this background, and their limitations.

Firstly, the triplet product could be generated by direct $S_1\to T_1$ intersystem crossing from the excited singlet state of PTZ--Ph$_n$--PDI,
\begin{equation}
^1\hbox{D-B-A}^* \ \longrightarrow\ ^3\hbox{D-B-A}^*, 
\end{equation}
or potentially (since the DFT calculations reported in the supplementary information of Ref.~\onlinecite{Weiss04} suggest that it is a borderline energetic possibility) by $S_0+S_1\to 2T_1$ singlet fission,
\begin{equation}
^1\hbox{D-B-A} +\ ^1\hbox{D-B-A}^*\ \longrightarrow\ 2\ ^3\hbox{D-B-A}^*, 
\end{equation}
before charge separation occurs. However, in a control experiment, Weiss {\em et al.} found a fluorescence quantum yield of 1 for model compounds of the form Ph$_n$--PDI, which seems to rule out any process competing with the fluorescence of $^1$PTZ--Ph$_n$--PDI$^*$ other than charge separation.\cite{Weiss04} (It is conceivable that the presence of the heavy sulphur atom in PTZ might facilitate $S_1\to T_1$ intersystem crossing, but this seems unlikely if the electronic excitation in $^1$PTZ--Ph$_n$--PDI$^*$ is confined to the PDI chromophore.)

\begin{figure}[t]
\centering
\resizebox{0.65\columnwidth}{!} {\includegraphics{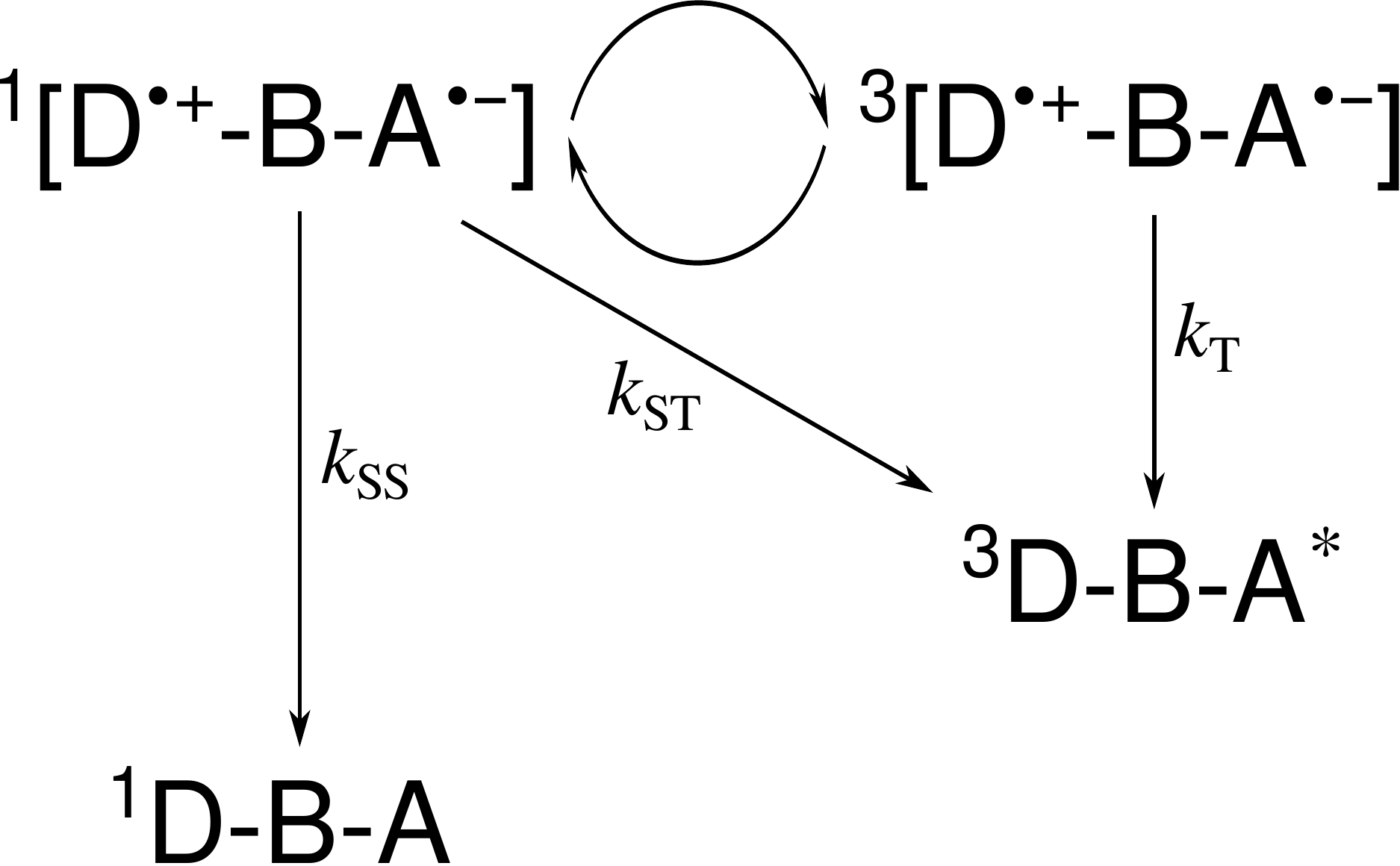}}
\caption{A recombination reaction scheme including the possibility of intersystem crossing accompanying charge recombination. Here, $k_{\rm SS}$ and $k_{\rm ST}$ are the rate constants for recombination of the singlet radical pair to the singlet and triplet product states respectively.}
\label{Updated_Scheme}
\end{figure}

Secondly, intersystem crossing accompanying charge recombination, 
\begin{equation}
^1\left[\hbox{D}^{\bullet+}\hbox{-B-A}^{\bullet-}\right] \ \longrightarrow\ ^3\hbox{D-B-A}^*,
\end{equation}
has been observed in these molecular wires.\cite{Dance06} This process is shown in the modified radical pair reaction scheme in Fig.~\ref{Updated_Scheme}, where it is labelled with the rate constant $k_{\rm ST}$ to distinguish it from `normal' recombination of the singlet radical pair to the ground state, now labelled $k_{\rm SS}$. Defining $f = k_{\rm ST} / (k_{\rm ST} + k_{\rm SS})$ as the fraction of the singlet radical pair which reacts to form the the triplet product, the total triplet yield from this scheme is
\begin{equation}
\begin{aligned}
\Phi^{\prime}_{\rm T}(B) & = \Phi_{\rm T}(B) + f\Phi_{\rm S}(B) \\
		   & = (1-f)\Phi_{\rm T}(B) + f,
\end{aligned}
\label{NewPhi}
\end{equation}
where $\Phi_{\rm T}$ is still defined by Eq.~\eqref{tyield}. Comparing $\Phi_{\rm T}'(B)/\Phi_{\rm T}'(0)$ with Eq.~\eqref{background}, it is clear that
\begin{equation}
x=\frac{f}{1-f},
\end{equation}
and therefore that this mechanism will have the same effect as adding a field-independent background contribution to the triplet yield. However, the analysis of the measurements in Ref.~\onlinecite{Dance06} found that the process in Eq.~(19) only occurs at temperatures below 200 K for PTZ--Ph$_n$--PDI wires with $n\ge 2$, whereas the experiments of Weiss {\em et al.}\cite{Weiss04} that we have compared with here were performed at room temperature.\cite{Weiss05}

Finally, the background could be due to a fraction of the radical pairs being formed in the triplet state, rather than the singlet state, during the initial charge separation: 
\begin{equation}
^1\hbox{D-B-A}^* \ \longrightarrow\ ^3\left[\hbox{D}^{\bullet+}\hbox{-B-A}^{\bullet-}\right].
\end{equation}
We have checked in our simulations that the triplet recombination rates in these wires are sufficiently fast that this would again be tantamount to adding a field-independent background contribution to the triplet yield of the radical pair recombination reaction. However, the fraction of radical pairs formed in the triplet state by this mechanism would have to be $f=x/(x+1)=0.04$ when $n = 2$ and 0.34 when $n = 3$ in order to explain our results. The first of these fractions is consistent with the amount of intersystem crossing observed during charge separation in other radical pair reactions,\cite{Maeda11} but the second is much larger than would be expected on the basis of previous experiments. While the intersystem crossing will be promoted by the spin-orbit coupling associated with the sulphur atom in the PTZ radical, we can see no reason why the fraction of radical pairs formed in the triplet state would increase upon changing the bridge length from 2 to 3 para-phenylene units.

In summary, while we believe that some field independent background contribution is required to explain the experimental triplet yields in Fig.~5, as discussed in Sec.~IV.A and justified \emph{a posteriori} in Sec.~V.C, we remain unconvinced by all of the possible mechanisms we have suggested to account for this. It would be interesting if further experiments could be done to shed more light on the processes in Eqs.~(17), (18), (19) and (22)  in an attempt to resolve this issue. Especially since the need for a background correction to explain the high field behaviour of the triplet yield does not seem to be confined to these particular PTZ$^{\bullet+}$-Ph$_n$-PDI$^{\bullet-}$ molecular wires: relative triplet yields $\Phi_{\rm T}(B)/\Phi_{\rm T}(0)$ that are significantly larger than 1/3 in the high field limit have also been observed by Wasielewski and co-workers for a variety of other molecular wires.\cite{Wasielewski06,Weiss03,Weiss05} \\

\section{Conclusions}

In this paper, we have used quantum spin dynamics simulations to reproduce the magnetic field effects on the triplet and radical pair yields measured by Weiss \emph{et al.}\cite{Weiss04} for a series of PTZ$^{\bullet+}$--Ph$_n$--PDI$^{\bullet-}$ molecular wires with increasing bridge lengths. We have extracted recombination rates for the singlet and triplet states of the radical pair from our simulations, and used these to shed light on the spin dynamics and charge recombination mechanisms of these molecular wires. The wide peaks in the triplet and radical pair yield MFEs observed experimentally and reproduced in our simulations are a result of the lifetime broadening of the triplet states of the radical pair. The triplet charge recombination rates follow a single exponential decay as a function of radical separation, consistent with the superexchange recombination mechanism. By contrast, the singlet rates are very similar for wires with bridges consisting of three or more para-phenylene units, suggesting that incoherent hopping is the primary recombination mechanism in the singlet pathway. The difference between the mechanisms of the two pathways can be explained using Marcus theory. And finally, we have found what we believe to be strong evidence for a magnetic field-independent background contribution to the triplet yield of the charge recombination reaction, the physical origin of which remains an open question.\\

\begin{acknowledgments}
We are grateful to Peter Hore for suggesting this application, and to both him and Michael Wasielewski for helpful discussions. Thomas Fay is supported by a Clarendon Scholarship from Oxford University and by the EPRSC Centre for Doctoral Training in Theory and Modelling in the Chemical Sciences, EPSRC grant no. EP/L015722/1. 
\end{acknowledgments}


\begin{thebibliography}{100}
\bibitem{Footnote} We found that the position of the minimum was independent of the rate constants $k_{\rm S}$ and $k_{\rm T}$.
\end{thebibliography}
\end{document}